\documentclass[prb,fleqn,12pt,showpacs]{revtex4-1}
\usepackage{epsfig}
\usepackage{graphicx}
\usepackage{amsfonts,amssymb}
\usepackage{amsmath}

\newcommand{\be}{\begin{equation}}
\newcommand{\ee}{\end{equation}}   
\newcommand{\bea}{\begin{eqnarray}}
\newcommand{\eea}{\end{eqnarray}}
\newcommand{\ba}{\begin{array}}
\newcommand{\ea}{\end{array}}

\newcommand{\phrl}[1]{Phys.~Rev.~Lett. {\bf #1}}
\newcommand{\phrb}[1]{Phys.~Rev.~B {\bf #1}}

\newcommand{\q}{{\bf q}}
\renewcommand{\k}{{\bf k}}

\begin{document}
\title{On the origin of CE-type orbital fluctuations in the ferromagnetic metallic La$_{1.2}$Sr$_{1.8}$Mn$_2$O$_7$}
\author{Dheeraj Kumar Singh$^1$$^,$$^2$}
\email{takimoto@hanyang.ac.kr}
\author{Ki Hoon Lee$^2$$^,$$^3$}
\author{Tetsuya Takimoto$^1$}
 
\affiliation{$^1$Department of Physics, Hanyang University, 
17 Haengdang, Seongdong, Seoul 133-791, Korea}
\affiliation{$^2$Asia Pacific Center for Theoretical Physics, Pohang, Gyeongbuk 790-784, Korea}
\affiliation{$^3$ Department of Physics, Pohang University of Science and Technology, Pohang, Gyeongbuk 790-784, Korea}

\begin{abstract}
We investigate the orbital fluctuations in the ferromagnetic-metallic phase of La$_{1.2}$Sr$_{1.8}$Mn$_2$O$_7$ by considering
a two orbital model within a tight-binding description which reproduces the ARPES Fermi surface. We find strong antisymmetric transverse orbital fluctuations at 
wavevector ($ 0.5 \pi, 0.5 \pi$) resulting from the Fermi-surface nesting
between the portions of bonding and antibonding bands instead of the widely believed nesting between the portions of bonding band despite 
their flat segments, which provide an insight into the origin of so called CE-type orbital fluctuations in the ferromagnetic-metallic phase.
Subsequent renormalization of the phonons near wavevector ($0.5 \pi,0.5 \pi$) and the behavior
of the phonon linewidth as a function of momentum are in agreement with the inelastic neutron scattering experiments. 

\end{abstract}
\maketitle
\newpage

\section{Introduction}
The vigorous competition among the multiplicity of phases resulting from the interactions of spin, charge,
orbital, and lattice degrees of freedom is central to the physics of manganites. The interplay of various degrees of freedom 
leads not only to the complex phase diagram but also to the unusual transport properties, especially the colossal magnetoresistance (CMR)
observed close to the ferromagnetic transition temperature.\cite{helmolt,jin} The CMR effect has been linked to the nano-scale 
charge and orbital correlations\cite{sen} which persist even in the ferromagnetic metallic phase.\cite{weber} Furthermore, Raman 
scattering experiment displays a sharp increase of the dynamical charge and orbital correlations near but above 
the ferromagnetic transition as a function of temperature, which is similar to the steep 
change in resistivity profile accompanying the transition.\cite{saitoh} The short-range charge and orbital correlations are
observed as diffuse peaks in the neutron scattering experiments\cite{adams,moussa_2007,weber} at wavevector (0.5$\pi$, 0.5$\pi$, 0), which is the same position 
where the super-lattice peaks are observed in the CE-type charge-orbital order.\cite{wollan_1955,goodenough_1955}     
 Such correlations, apart from playing a crucial role in the CMR effect, renders the ferromagnetic metallic
 phase disparate from an ideal ferromagnet. For instance, the spin-wave measurements exhibit several
 anomalies including the softening near zone boundary in the $\Gamma$-X 
direction,\cite{hwang,ye} wherein the role of orbital correlations has been emphasized with dominant contribution to
the magnon self-energy coming from the orbital-fluctuation modes near the wavevector (0.5$\pi$, 0.5$\pi$, 0).\cite{dk1,dk2} 

Layered La$_{2-2x}$Sr$_{1+2x}$Mn$_2$O$_7$ belongs to the Ruddlesten-Popper series with n=2, where bilayers of MnO$_6$ 
octahedra are separated by (La, Sr) O. The transport and magnetic properties exhibit high anisotropy due to 
the tetragonal crystal structure while the reduced dimension further 
boosts the CMR effect as in the case of La$_{1.2}$Sr$_{1.8}$Mn$_2$O$_7$. The transition from
paramagnetic insulating to ferromagnetic metallic phase with the in-plane Mn spins having a
saturation moment $\approx$ 3 $\mu_B$/Mn appears at $\approx$ 120$K$, although the ferromagnetic phase
displays a large resistivity in contrast with a normal metal.\cite{moritomo,mitchell}

Recent angle resolved photoemission spectroscopy (ARPES) measurements on the Fermi surface of layered manganites have provided crucial insight into the formation of 
both long-range and nano-scale charge-orbital structures.\cite{chuang,mannela,evtushinsky} Low temperature ARPES data on the ferromagnetic-metallic La$_{1.2}$Sr$_{1.8}$Mn$_2$O$_7$ consists of multiple Fermi surfaces, a square-like 
 electron pocket around the $\Gamma$ point and two hole pockets around the M point corresponding to the bonding and antibonding bands due 
 to the bilayer lattice structure.\cite{zsun1,jong} Based on the shape of the Fermi surfaces, the nesting between the Fermi surfaces has 
been suggested to be responsible for the formation of short-range charge/orbital correlations. 
This is indeed also supported by the ARPES experiments which found pseudo-gap for the Fermi surface.\cite{chuang,mannela} 
On the other hand, quasi-particle peaks observed in several experiments subsequently\cite{zsun1,zsun2,jong} has been
attributed to the intergrowth present in the experimental samples by a recent scanning tunneling microscopy (STM) combined with ARPES, thus reaffirming
the pseudo-gap structure associated with the Fermi surface.\cite{masse}

The persistent CE-type dynamical short-range charge-orbital correlations in the ferromagnetic metallic phase at $x = 0.4$, apart 
from being responsible for the pseudo-gap structure in the Fermi surface, yields a strong renormalization of phonons near the wavevector 
(0.5$\pi$, 0.5$\pi$, 0) observed in the inelastic neutron scattering,\cite{weber} which is believed to be resulting from the 
strong Fermi surface nesting in the ferromangetic metallic state.
As the hole pocket corresponding to the bonding portion of the Fermi surface has nearly straight segments in
 comparison to the antibonding portions, bonding-bonding nesting has been suggested\cite{chuang,zsun3} to 
be responsible for the nano-scale charge-orbital correlations in the ferromagnetic metallic phase. However, there is an apparent discrepancy in the magnitude of the 
nesting wavevector $\approx$ (0.6$\pi$, 0.6$\pi$, 0) for the portions of bonding band and the wavevector (0.5$\pi$, 0.5$\pi$, 0) for 
the diffuse peak in the neutron scattering experiments. 

In this paper, we explore the orbital fluctuations in the ferromagnetic-metallic La$_{1.2}$Sr$_{1.8}$ \newline Mn$_2$O$_7$ by considering a model 
which captures the important characteristics of the ARPES Fermi surfaces. After identifying the dominant orbital fluctuation modes 
among several possible modes due to the multiple Fermi surfaces, our study of the impact of relevant orbital 
fluctuations on the phonons by calculating phonon self-energy, which can be observed in the inelastic neutron scattering experiments, provides 
the important link between the ARPES Fermi surface and dynamical orbital correlations observed in the neutron scattering experiments.
\section{Model Hamiltonian}
To study the orbital fluctuations of the ferromagnetic-metallic La$_{1.2}$Sr$_{1.8}$Mn$_2$O$_7$ with 
$e_g$ electron density $1 - x = 0.6$ per site, we consider an effective Hamiltonian which 
treats the large Hund's coupling ($J \sim W$)\cite{dagotto} of $e_g$ spin to $t_{2g}$ spins and
intra-orbital Coulomb interaction ($U$) for $e_g$ electrons at meanfield level. Although, the experimental saturation
moment 3 $\mu_B$/Mn differs slightly from total magnetic moment of electron number 3.6 including $e_g$ and $t_{2g}$ electrons, our assumption of completely 
empty minority-spin band
($\downarrow$-spin) is still reasonable in the case of large Hund's coupling. Then, the effective 
Hamiltonian spanned by $d_{x^2-y^2}$ and $d_{3z^2-r^2}$ orbitals on a bilayer lattice system is 
\bea
{\mathcal H}  &=& -\sum_{\gamma \gamma'\sigma  p p^{\prime} {\bf ia}}
  t^{p p^{\prime}{\bf a}}_{\gamma \gamma'} d_{ {\bf i} p \gamma  \sigma }^{\dag}
  d_{{\bf i+a} p^{\prime} \gamma' \sigma   } +\Delta \sum_{p {\bf i}}\mathcal{T}^p_{{\bf i} z}
  -\sum_{{\bf i} p \gamma } \sigma (Um_{{\bf i} p\gamma}/2+JS_{{\bf i} p})d_{ {\bf i} p \gamma  \sigma }^{\dag}
  d_{{\bf i} p \gamma \sigma } \nonumber \\
 &+& U'\sum_{{\bf i} p } n_{{\bf i} p \gamma}n_{ {\bf i} p \gamma'}+\sum_{ {\bf i} p}\sum_{ l=x,z }
 \omega_l f^{\dag}_{ {\bf i} p l}f_{{\bf i} p l} + \sum_{ {\bf i} p}\sum_{ l=x,z } 
 g (f_{{\bf i} p l}+f^{\dag}_{{\bf i} p l})\mathcal{T}^p_{{\bf i} l}.
\eea
Here, first term describes the nearest-neighbor electron transfer, where 
$d_{{\bf i} p 1 \sigma }^{\dag}$ ($d_{{\bf i} p 2  \sigma }^{\dag}$) is the electron 
creation operator for the orbital $d_{x^2-y^2}$ ($d_{3z^2-r^2}$) with spin $\sigma$ of plane $p$ in a unit cell
${\bf i}$. $t^{p p^{\prime}{\bf a}}_{\gamma \gamma'}$ are the
 nearest-neighbor hopping elements between $\gamma$ orbital of $p$ plane and $\gamma'$ orbital of $p^{\prime}$ plane along a two dimensional vector ${\bf a}$, which are given by
 $t^{pp {\bf x}}_{11}$ = $-\sqrt{3}t^{pp {\bf x}}_{12}$
 = $-\sqrt{3}t^{pp {\bf x}}_{21}$ = $3t^{pp {\bf x}}_{22}$ = $3t_\parallel/4$ and $t^{pp{\bf y}}_{11}$
 = $\sqrt{3}t^{pp{\bf y}}_{12}$ = $\sqrt{3}t^{pp{\bf y}}_{21}$ = 
$3t^{pp{\bf y}}_{22}$ = $3t_\parallel/4$ for the inplane hopping for each plane, and 
$t^{p\bar{p}{\bf 0}}_{11}$ = $t^{p\bar{p}{\bf 0}}_{12}$ = $t^{p\bar{p}{\bf 0}}_{21}$ = 0 and $t^{p\bar{p}{\bf 0}}_{22}$ = $t_{\perp}$ 
for the hopping between two planes.  Crystal-field splitting of e$_g$ levels 
in the tetragonal symmetry is taken into account by the second term. The third term describes the total ferromagnetic exchange term arising due to the Hund's coupling 
 and intra-orbital Coulomb interaction in the ferromagnetic phase, where $m_{{\bf i} p}$ and $S_i$ are magnetization for e$_g$ electrons and $t_{2g}$ spin, 
 respectively. Fourth term is the inter-orbital Coulomb
 interaction term, where Hund's coupling and pair-hopping terms for the $e_g$ orbitals are ignored for simplicity. The fifth term describes the local Jahn-Teller phonons with $f_{{\bf i} p x }^{\dag}$
and $f_{{\bf i} p z }^{\dag}$ as the phonon creation operators for the transverse $({x^2-y^2})$- and longitudinal
$({3z^2-r^2})$-type distortions, respectively. Sixth term represents the  
electron-phonon coupling with the transverse and longitudinal orbital operators given by $\mathcal{T}^p_{{\bf i} x }$ and $\mathcal{T}^p_{{\bf i} z }$, 
respectively. Herefrom, spin index $\sigma$ is dropped based on the assumption that only majority spin
band is occupied. Then, the longitudinal and transverse orbital operators 
are given by $\mathcal{T}^p_{ {\bf i} z} = 
 \psi^\dagger_ { {\bf i} p } \hat{\tau}_z \psi_{{\bf i} p }$ 
 and $\mathcal{T}^p_{{\bf i} x } =  \psi^\dagger_ {{\bf i} p }\hat{\tau}_x \psi_{{\bf i} p  }$ with  $\psi^\dagger_ {{\bf i} p  } = 
(d^\dagger_{{\bf i} p 1 }, d^\dagger_{{\bf i} p 2})$, where $\hat{\tau}_z$ and $\hat{\tau}_x$ are the $z$- and $x$-components of Pauli matrices in the
orbital space, respectively.
 We assume $\omega_x$ = $\omega_z$ for simplicity even though our system has tetragonal symmetry. 

In the following, it is convenient to introduce the symmetric and antisymmetric 
operators for both the electron and phonon operators with respect to the mirror symmetry between two planes in the 
unit cell,\cite{eremin}
\be
d_{{\bf i} s \gamma } =\frac{1}{\sqrt{2}}(d_{{\bf i} 1 \gamma } + d_{{\bf i} 2 \gamma  }), \,\,
d_{{\bf i} a \gamma } = \frac{1}{\sqrt{2}}(d_{{\bf i} 1 \gamma \sigma} - d_{{\bf i} 2 \gamma }),
\ee
and 
\be
f_{{\bf i} s l}  = \frac{1}{\sqrt{2}}(f_{ {\bf i} 1 l} + f_{{\bf i} 2 l}), \,\, f_{ {\bf i} a l} = \frac{1}{\sqrt{2}}(f_{ {\bf i} 1 l} - f_{{\bf i} 2 l}), 
\ee
which leads to
\bea
 d_{{\bf i} p \gamma  } & = & \frac{1}{\sqrt{2}}(d_{{\bf i} s 
\gamma }
+ (-1)^{p-1} d_{{\bf i} a \gamma }), \nonumber\\ 
f_{{\bf i} p l}  &=&  \frac{1}{\sqrt{2}}(f_{ {\bf i} s l} + (-1)^{p-1}f_{{\bf i} a l}).
\eea
Then, the orbital operators can be expressed using these symmetric and antisymmetric operators as
\be
\mathcal{T}^{p}_{{\bf i} l}  =  \frac{1}{2}\sum_{\gamma \gamma^{\prime} } \tau^{l}_{\gamma \gamma^{\prime}}[d_{ {\bf i} s \gamma   }^{\dag}
  d_{{\bf i} s \gamma^{\prime}  }+d_{ {\bf i} a \gamma   }^{\dag}
  d_{{\bf i} a \gamma^{\prime}  }+(-1)^{p-1}(d_{ {\bf i} s \gamma  }^{\dag}
  d_{{\bf i} a \gamma^{\prime}  }+d_{ {\bf i} a \gamma   }^{\dag}
  d_{{\bf i} s \gamma^{\prime} \ })] .
\ee

Therefore, the coupling term for the electron and the transverse phonon in the Hamiltonian is transformed to 
\bea
\sum_{{\bf i}}\frac{g}{\sqrt{2}}[(f_{{\bf i} s x } + f^{\dagger}_{ {\bf i}  s x})(d^{\dagger}_{ {\bf i} s 1  } d_{ {\bf i} s 2 }+d^{\dagger}_{{\bf i} a 1 } d_{{\bf i} a 2 
 }+
d^{\dagger}_{{\bf i} s 2  } d_{ {\bf i} s 1 }+d^{\dagger}_{ {\bf i} a 2 } d_{ {\bf i} a 1 })\nonumber\\
+(f_{{\bf i} a x } + f^{\dagger}_{{\bf i} a x})(d^{\dagger}_{ {\bf i} s 1 } d_{ {\bf i} a 2  }+d^{\dagger}_{{\bf i} a 1  }
d_{{\bf i} s 2  }+
d^{\dagger}_{{\bf i} s 2  } d_{{\bf i} a 1  }+d^{\dagger}_{{\bf i} a 2 } d_{ {\bf i} s 1 })],
\eea
and similarly the coupling term for the electron and the longitudinal phonon to
\bea
\sum_{{\bf i}}\frac{g}{\sqrt{2}}[(f_{ {\bf i} s z} + f^{\dagger}_{ {\bf i} s z})(d^{\dagger}_{ {\bf i} s 1 } d_{{\bf i} s 1  }+d^{\dagger}_
{ {\bf i} a 1 } d_{ {\bf i}  a 1 }-
d^{\dagger}_{ {\bf i} s 2  } d_{{\bf i} s 2 }-d^{\dagger}_{{\bf i} a 2} d_{{\bf i}  a 2 })\nonumber\\
+(f_{ {\bf i} a z} + f^{\dagger}_{{\bf i} a z})(d^{\dagger}_{ {\bf i} s 1 } d_{{\bf i} a 1 }+d^{\dagger}_{{\bf i} a 1  } d_{{\bf i} s 1 }-
d^{\dagger}_{{\bf i} s 2 } d_{{\bf i} a 2 }-d^{\dagger}_{{\bf i} a 2  } d_{{\bf i} s 2 })].
\eea

Using the symmetric and antisymmetric operators, the kinetic and CEF terms of Eq. (1) can be expressed in the block-diagonal form after the Fourier
transformation as
\be
H_{\rm kin}({\bf k})+H_{\rm CEF}=  \sum_{{\bf k},\gamma} \psi_{{\bf k}} ^{ \dagger} 
\begin{pmatrix} 
 \hat{H}^{ss}({\bf k}) & \hat{0} \\
 \hat{0} & \hat{H}^{aa}({\bf k})
\end{pmatrix} \psi_{{\bf k} },
\ee
where $\psi^{\dag}_{{\bf k} }=(d^{\dag}_{s 1  }({\bf k}), d^{\dag}_{s 2  }({\bf k}), d^{\dag}_{a 1  }
({\bf k}), d^{\dag}_{a 2 }({\bf k}))$. $\hat{0}$ is a 2$\times$2 matrix with all vanishing elements,
and $\hat{H}^{ss(aa)}({\bf k})$ is a 2$\times$2 matrix given by 
\bea
 \hat{H}^{ss(aa)}({\bf k}) &=& 
\left(\varepsilon^{s(a)}_{+ }({\bf k})-\mu\right) \hat{\tau}_0 +\varepsilon^{s(a)}_{-}({\bf k})
\hat{\tau}_z+\varepsilon_{12}({\bf k})\hat{\tau}_x 
\eea
with $\mu$ as the chemical potential. $\hat{\tau}_0$ is the unit matrix in the orbital space. The coefficients of the unit and the Pauli matrices are given by 
\bea
  \varepsilon^{s(a)}_+({\bf k}) & = & \frac{1}{2}t_\parallel (\varepsilon_1({\bf k})+\varepsilon^{s(a)}_2({\bf k})) \nonumber\\ 
  \varepsilon^{s(a)}_-({\bf k}) & = &
  \frac{1}{2}t_\parallel(\varepsilon_1({\bf k})-
  \varepsilon^{s(a)}_2({\bf k}))+\Delta \nonumber\\
     \varepsilon^{12}({\bf k}) & = & \frac{\sqrt{3}}{2}t_\parallel(\cos k_x-\cos k_y)
\eea
with
\bea
    \varepsilon_1({\bf k}) & = & -\frac{3}{2}t_\parallel(\cos k_x+\cos k_y)\nonumber\\
    \varepsilon^{s(a)}_2({\bf k}) & = &  -\frac{1}{2}t_\parallel(\cos k_x+\cos k_y)\mp t_{\perp}. 
\eea
Here, $\epsilon^s_2 ({\bf k})$ ($\epsilon^a_2 ({\bf k})$) has negative (positive) sign in front of $t_{\perp}$. Single electron Matsubara Green's
function is described in the block-diagonal 
matrix form
\be
\hat{G}^{} ({\bf k}, i \omega_n ) = 
\begin{pmatrix} 
 \hat{G}^{ss}({\bf k}) & \hat{0} \\
 \hat{0} & \hat{G}^{aa}({\bf k})
\end{pmatrix},
\ee
where 
\be
\hat{G}^{ss(aa)} ({\bf k}, i \omega_n ) = \frac{\left( i \omega_n - \varepsilon^{s(a)}_{+}({\bf k}) +\mu \right)
\hat{\tau}_0 - \varepsilon^{s(a)}_{-}({\bf k}) 
\hat{\tau}_z + \varepsilon_{12} ({\bf k}) \hat{\tau}_x}{\left(i \omega_n - E^{s(a)}_{+}({\bf k}) \right) \left(i \omega_n -
E^{s(a)}_{-}({\bf k}) \right)},
\ee

\be
E^{s(a)}_{\pm }({\bf k}) =  \varepsilon^{s(a)}_{+}({\bf k}) \pm \sqrt{(\varepsilon^{s(a)}_{-}({\bf k}))^2 +
(\varepsilon_{12}({\bf k}))^2} - \mu\label{Ek}, 
\ee
and Fermionic Matsubara frequency $\omega_n=(2n+1)\pi T$. Here, we note that there are 
two bands corresponding to each of the symmetric and antisymmetric Green's functions.

\begin{figure}
\begin{center}
\vspace*{-2mm}
\hspace*{-0mm}
\psfig{figure=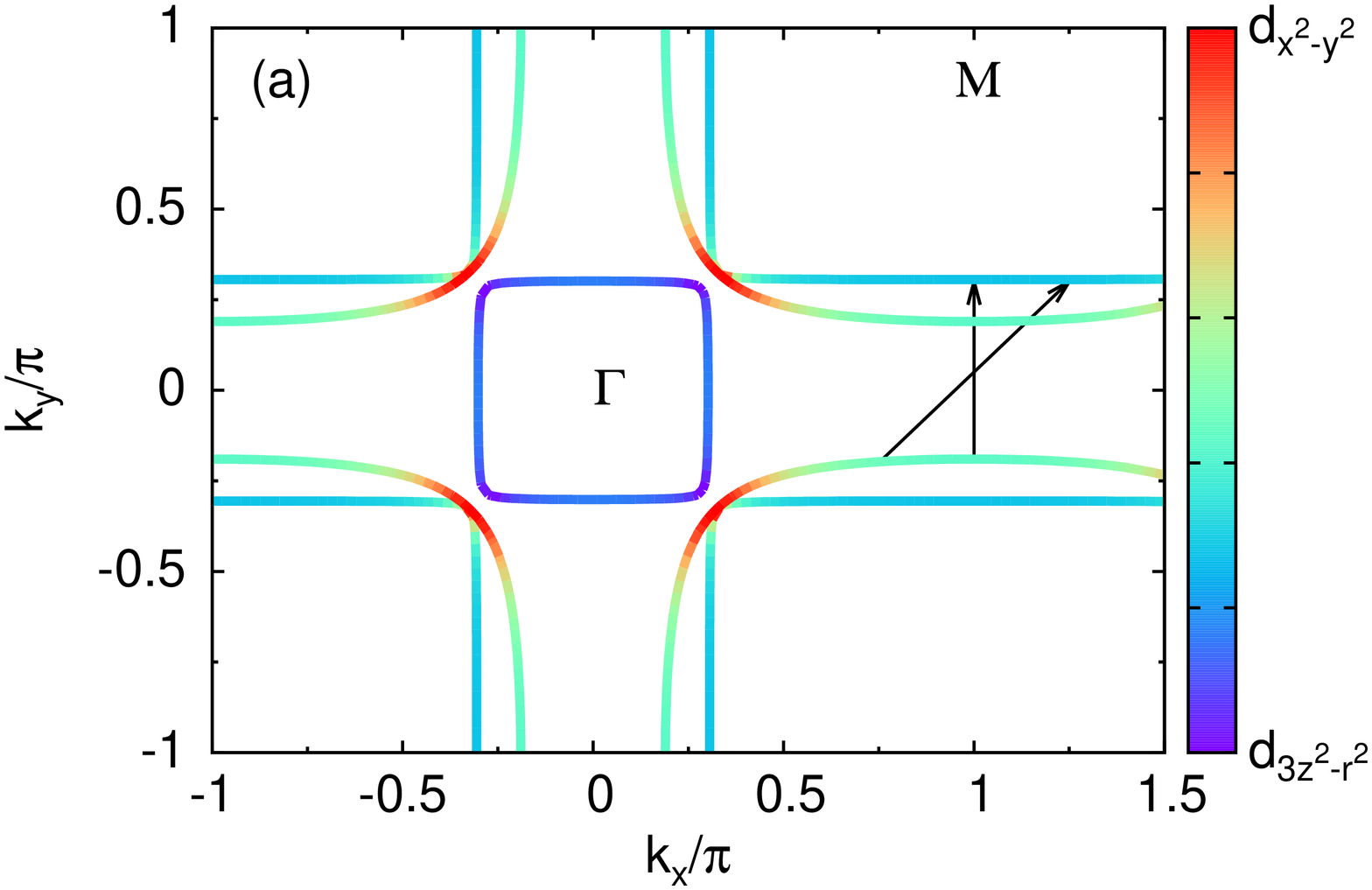,width=55mm,angle=-90}
\hspace*{5mm}
\psfig{figure=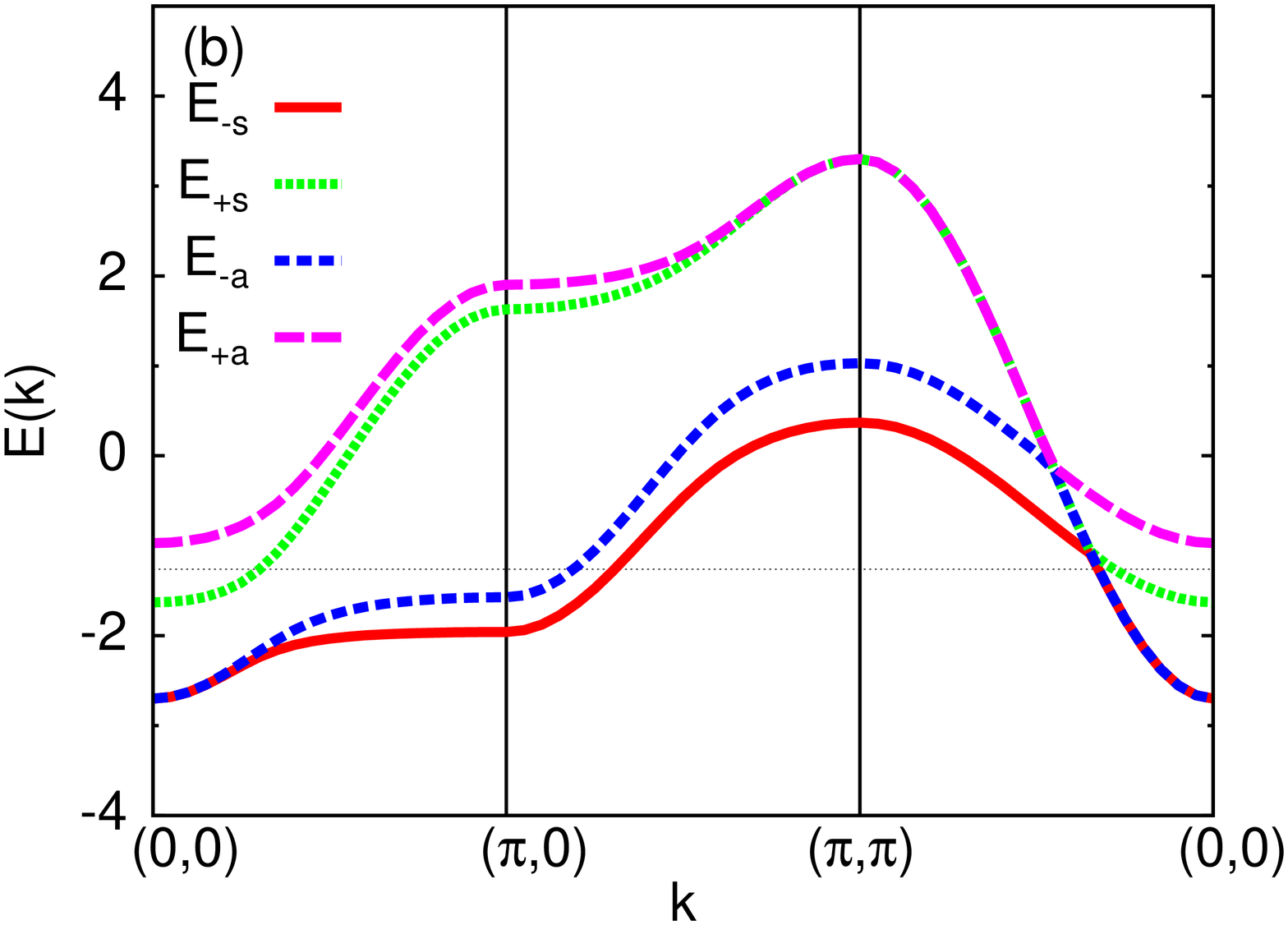,width=49mm,angle=-90}
\vspace*{-10mm}
\end{center}
\caption{(a) Calculated Fermi surfaces for La$_{1.2}$Sr$_{1.8}$Mn$_2$O$_7$ with $d_{x^2-y^2}$ and $d_{3z^2-r^2}$ orbital densities, and (b) electronic dispersion in 
the high-symmetry directions for the crystal-field parameter $\Delta = -0.3$,  the chemical potentials $\mu = -1.26$, and the inter-planar
 hopping $t_{\perp} = 0.33$.}
\label{fermis}
\end{figure}  
To reproduce the experimental Fermi surfaces with the above dispersion, we choose the values of chemical
potential $\mu$ = -1.26, inter-planar hopping parameter $t_{\perp} = 0.33$, 
and crystal-field parameter $\Delta = 0.3$ in the unit of $t_\parallel = t$. The chemical
potential $\mu$ = -1.26 corresponds to the $e_g$ electron density of $1-x \approx 0.54$/Mn, inter-planar hopping controls 
the splitting of the hole pockets around X, and 
crystal-field parameter improves the nesting quality by straightening the hole pockets as shown in Fig. 1(a). The
electron pocket around the $\Gamma$ point and two hole pockets around the M point with one having almost 
straight segments belonging to the lower of plane-symmetric bands (bonding band) are in good agreement with the ARPES measurements.\cite{zsun4} The electron 
pocket has predominantly $d_{3z^2-r^2}$ orbital character whose proportion decreases only slightly on moving 
from $\Gamma$-M to $\Gamma$-X direction. Both the hole pockets have large $d_{x^2-y^2}$ orbital character in the 
$\Gamma$-M direction. In addition, the hole pocket 
belonging to the lower of plane-antisymmetric band (antibonding band) contains an almost equal mixture of 
$d_{x^2-y^2}$ and $d_{3z^2-r^2}$ orbitals near X at the zone boundary, whereas the hole pocket 
belonging to the bonding band consists mainly of $d_{3z^2-r^2}$ orbital. Similar features were also obtained in the band-structure
calculations.\cite{saniz} Particularly along the $\Gamma$-M direction, since the orbital 
mixing vanishes, the electronic states on the electron and hole Fermi surface 
have the orbital characters of $d_{3z^2-r^2}$ and $d_{x^2-y^2}$, respectively. There exist both intra-band (bonding-bonding or 
antibonding-antibonding) and inter-band (bonding-antibonding) nestings. From Fig. 1(a), the nesting vectors for the bonding-bonding,
antibonding-antibonding, and bonding-antibonding nestings are expected at $\{(0.6 \pi, 0),(0.6 \pi, 0.6 \pi) \}$, $
\{(0.4 \pi, 0),(0.4 \pi, 0.4 \pi) \}$, and $\{(0.5 \pi, 0),(0.5 \pi, 0.5 \pi)\}$, respectively. The quality of nesting decreases from 
the bonding-bonding to the bonding-antibonding case, and is the poorest for antibonding-antibonding nesting. 

\section{Orbital Fluctuation}

To identify the important low energy orbital excitations in the ferromagnetic-metallic phase of 
the bilayer manganites with the spin degree of freedom already frozen, we 
investigate the orbital susceptibility of a bilayer lattice system with two orbitals at each site, which, in general, will be a 16$\times$16 
matrix as 16 different orbital operators can be defined due to the four different electron field
operators corresponding to the orbital and planar degrees of freedom in a unit cell. But the relevant orbital 
susceptibility, which involves only the $intralayer$ orbital operators, can be written in a 8$\times$8 matrix
form. A further simplification can be achieved by changing to a basis in the planar space, which involves orbital 
operators symmetric and 
antisymmetric with respect to the mirror symmetry between the two planes in a unit cell. The relevant orbital susceptibility reduces to two 4$\times$4 matrices
in this basis resulting from the fact that the correlation functions involving symmetric and antisymmetric orbital operators 
vanish identically by the symmetry of the bilayer Hamiltonian, and therefore can be defined as follows:
\be
\chi^{m}_{ll^{\prime}}(\q,i\Omega_n)= 2\int^{\beta}_0{d\tau e^{i \Omega_{n}\tau}\langle T_\tau
[{\cal T}^{m}_{\q l}(\tau) {\cal T}^{m}_{ -\q l^{\prime}}(0)]\rangle}.
\ee
Here, $\langle...\rangle$ denotes thermal average, $T_\tau$ imaginary time ordering, and $\Omega_n$ are the Bosonic
Matsubara frequencies. ${\cal T}^m_{\q l}$ is the Fourier $\q$-component of orbital operator ${\cal T}^m_{{\bf i} l}$ symmetric ($m = s$) 
and antisymmetric ($m = a$) with respect to the mirror symmetry between the two planes in a unit cell. The symmetric and antisymmetric components are 
given by 
\be
{\cal T}^s_{{\bf q} l} = ({\cal T}^1_{{\bf q}l} +{\cal T}^2_{{\bf q}l})/2, \,\,\,\,\,
{\cal T}^a_{ {\bf q} l} = ({\cal T}^1_{{\bf q}l} -{\cal T}^2_{{\bf q}l})/2,
\ee
respectively. 

Expression for the relevant RPA-level orbital susceptibility is given by\cite{takimoto}
\bea
    && \hat{\chi}^{s(a)}({\bf q}, i \Omega_n) \!=\!\hat{\chi}^{0;s(a)}({\bf q}, i \Omega_n)
  [\hat{1}+\hat{U}^{}\hat{\chi}^{0;s(a)}({\bf q}, i \Omega_n)]^{-1}.
  \eea
After a unitary transformation, row and column labels appear in the order 11, 22, 12, and 21 with 1 and 2 as orbital indices. 
$\hat{1}$ is a 4$\times$4 matrix. The matrix elements of $\hat{\chi}^{0;s(a)}({\bf q}, i \Omega_n)$ are defined as 
\bea
\chi^{0;s}_{\mu\nu,\alpha\beta}({\bf q}, i \Omega_n)&=&(\chi^{0;aa,aa}_{\mu\nu,\alpha\beta}({\bf q}, i \Omega_n)
+\chi^{0;ss,ss}_{\mu\nu,\alpha\beta}({\bf q}, i \Omega_n))/2\nonumber\\
\chi^{0;a}_{\mu\nu,\alpha\beta}({\bf q}, i \Omega_n)&=&(\chi^{0;as,as}_{\mu\nu,\alpha\beta}({\bf q}, i \Omega_n)
+\chi^{0;sa,sa}_{\mu\nu,\alpha\beta}({\bf q}, i \Omega_n))/2,
\eea
where $\chi^{0;rs,uv}_{\mu\nu,\alpha\beta}({\bf q}, i \Omega_n)$=
$-T\sum_{{\bf k},m}
G^{ru}_{\alpha\mu}({\bf k}+{\bf q},i\omega_{m}+i\Omega_n)
G^{sv}_{\nu\beta}({\bf k},i\omega_{m})$, and $G^{sv}_{\nu\beta}({\bf k},i\omega_{n})$ is block diagonalized with respect to the superscripts ($s$ and $v$) as already
discussed in 
the previous section. The bare symmetric and antisymmetric susceptibilities can be expressed explicitly as follows 
\begin{eqnarray}
\chi^{0;s}_{\mu \nu, \alpha \beta}(\q, i \Omega_n) =-\frac{1}{N}\sum_{\k } \sum_{i,j} \sum_{ \pm} \sum_{p = s,a }
a^{\mu p^*}_{j\k + \q} a^{\nu p}_{i \k} a^{\beta p^* }_{i \k} a^{\alpha p}_{j
\k + \q}\frac { n(E^p_{i}(\k))-n(E^p_{j}(\k+\q)) }{i \Omega_n+E^p_{i}(\k)-E^p_{j}(\k+\q) } \nonumber\\
\chi^{0;a}_{\mu \nu, \alpha \beta}(\q, i \Omega_n) =-\frac{1}{N}\sum_{\k } \sum_{i,j} \sum_{ \pm} \sum_{p = s,a }
a^{\mu \bar{p}^*}_{j\k + \q} a^{\nu p}_{i \k} a^{\beta p^*}_{i \k} a^{\alpha \bar{p}}_{j \k + \q}\frac 
{ n(E^p_{i}(\k))-n(E^{\bar{p}}_{j}(\k+\q)) }{i \Omega_n+E^p_{i}(\k)-E^{\bar{p}}_{j}(\k+\q) },
\end{eqnarray}
where $a^{\nu p}_{i}$s are the unitary coefficients of the $i$-band to $\nu$-orbital obtained from Eq. (13), and $n(\xi)$ is the Fermi distribution function. Finally, the interaction matrix 
is given by 
\be
\hat{U}= \left\{
\ba{@{\,} l @{\,} c}
-U & (\mu=\alpha \ne \nu=\beta)\\
U & (\mu=\nu \ne \alpha = \beta)\\
0 & (\mathrm{otherwise})
\ea \right. ,
\label{eq_U}
\ee 
where the parameter $U$ has been used for the inter-orbital Coulomb interaction from now on.
\begin{figure}
\begin{center}
\vspace*{-2mm}
\hspace*{0mm}
\psfig{figure=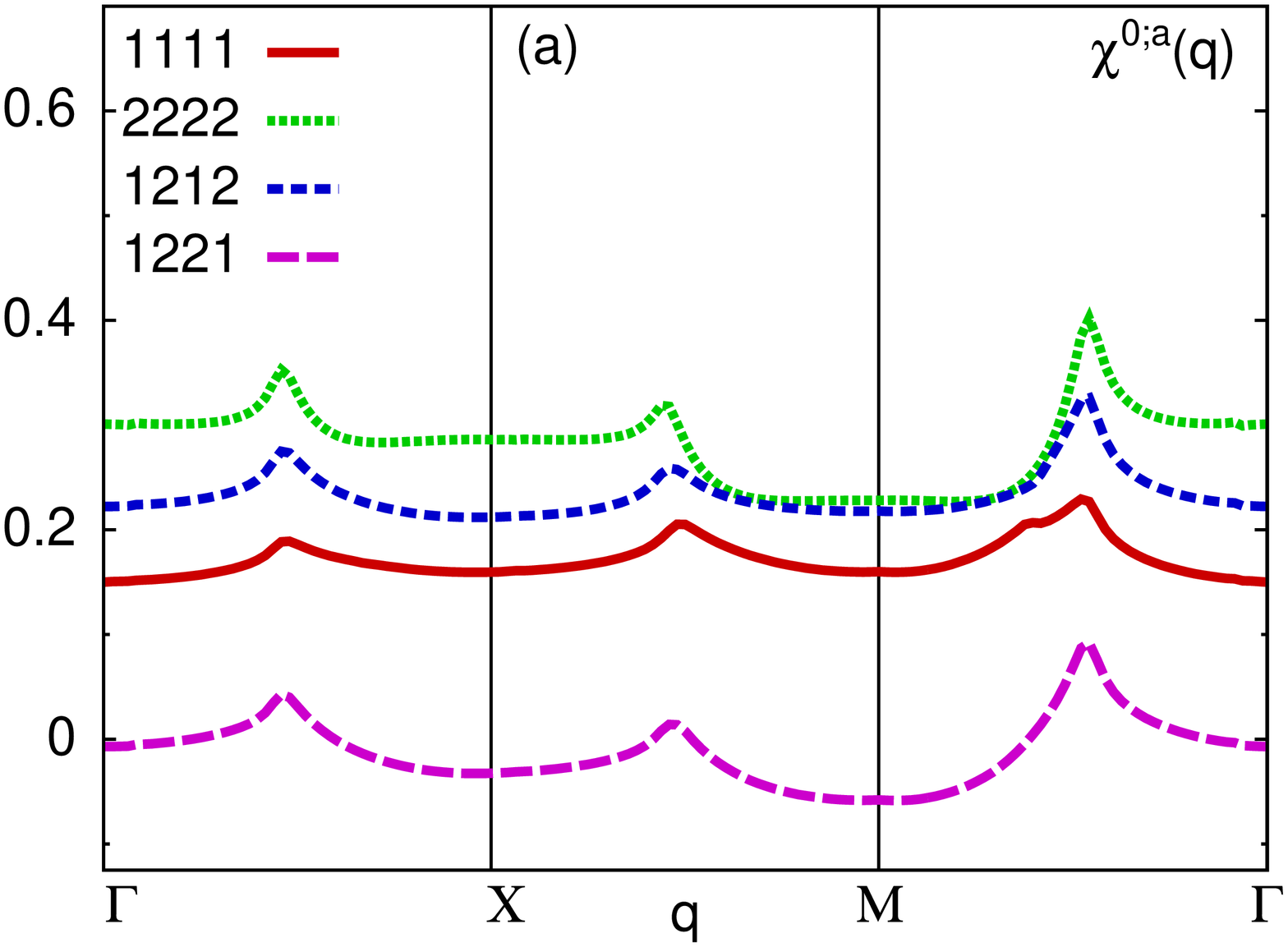,width=55mm,angle=-90}
\psfig{figure=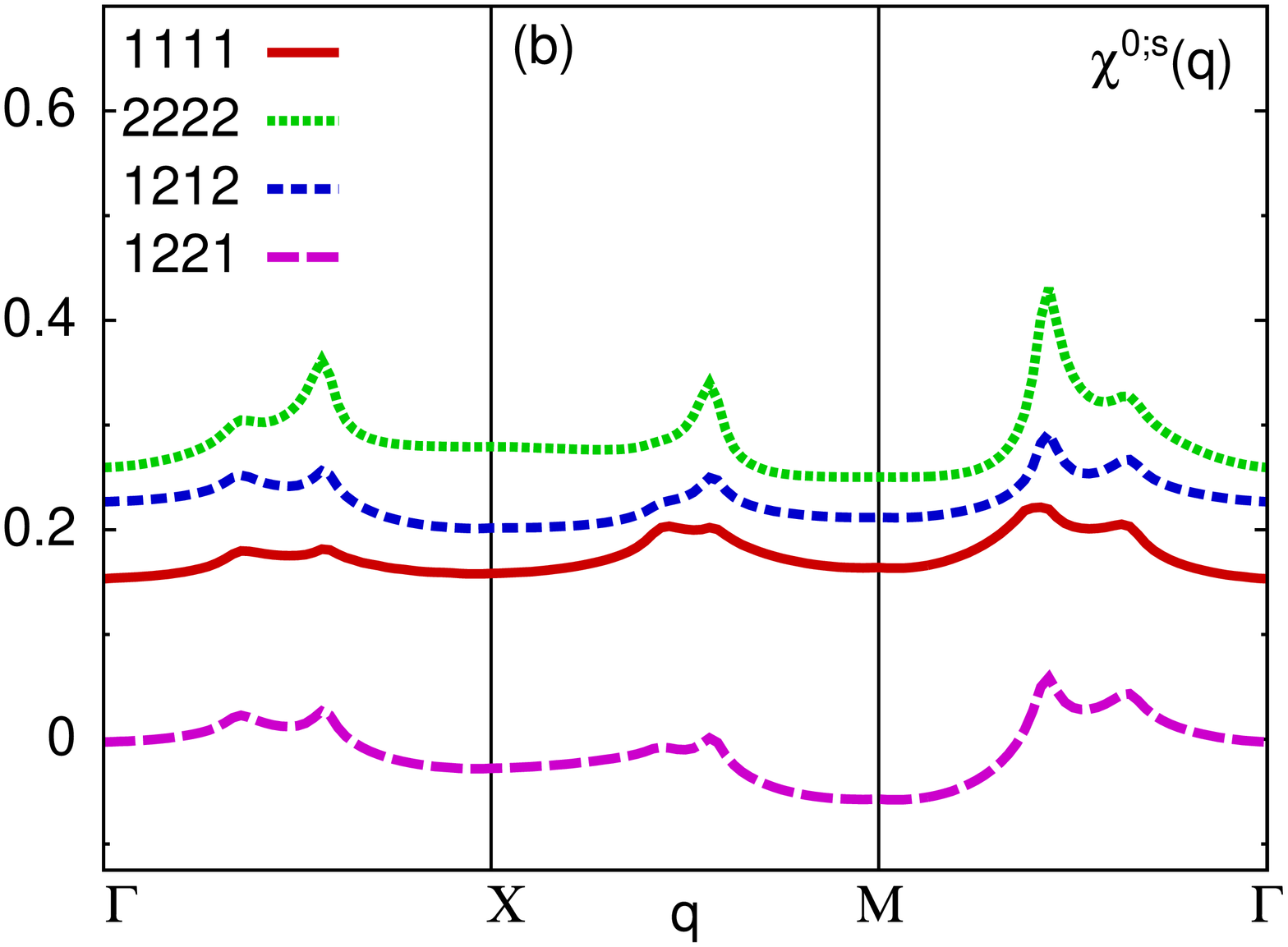,width=55mm,angle=-90}
\vspace*{-5mm}
\end{center}
\caption{Principal components contributing to the transverse and longitudinal susceptibility for the (a) antisymmetric and (b) symmetric case. 
The antisymmetric and symmetric susceptibility show peaks at $\{(0.5 \pi, 0)$, $(0.5 \pi, 0.5 \pi) \}$ and 
$\{(0.6 \pi, 0)$, $(0.6 \pi, 0.6 \pi)\}$, respectively.}
\label{bsus}
\end{figure}  

\begin{figure}
\begin{center}
\vspace*{-2mm}
\hspace*{0mm}
\psfig{figure=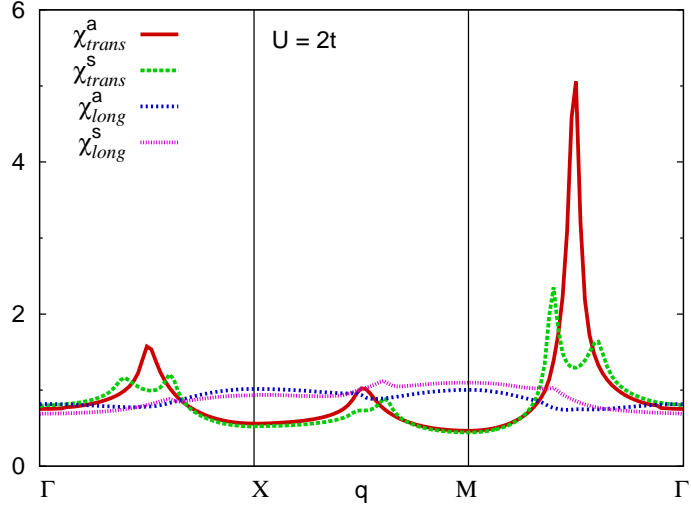,width=70mm,angle=-90}
\vspace*{-5mm}
\end{center}
\caption{RPA level transverse and longitudinal susceptibilities for symmetric and antisymmetric case  $U = 2.0t$. Electronic correlation 
enhances the the antisymmetric transversal orbital susceptibility at $(0.5 \pi, 0.5 \pi$) more in comparison to others.}
\label{dens}
\end{figure} 
Fig. \ref{bsus} shows the principal components of one bubble static orbital susceptibility for the chemical potential 
$\mu$ = -1.26, interplane hopping parameter $t_{\perp}$ = 0.33, and temperature $T = 0.02$. The components of antisymmetric susceptibility 
exhibit peak structures at $\{(0.5 \pi, 0)$, $(0.5 \pi, 0.5 \pi) \}$ and those of symmetric 
show peaks at $\{(0.6 \pi, 0)$, $(0.6 \pi, 0.6 \pi)\}$ as shown in Fig. 2(a) and 2(b), respectively.  In the case of $\chi^{0;}_{1212}(\q)$ 
and $\chi^{0;}_{1221}(\q)$ contributing to the transverse 
orbital susceptibility, the antisymmetric components display a sharper and relatively larger peaks at $(0.5 \pi, 0.5 \pi)$ in comparison to the 
symmetric components at $(0.6 \pi, 0.6 \pi)$, and viceversa for the components $\chi^{0;}_{1111}(\q)$ and $\chi^{0;}_{2222}(\q)$
contributing to the charge or longitudinal orbital susceptibility. The above features are sensitive to both the orbital
composition and the quality of nesting. For instance, $\chi^{0;s}_{2222}(\q)$ shows largest peaks at $(0.6 \pi, 0.6 \pi)$ reflecting the straight segments of 
the Fermi surfaces dominated by $d_{3z^2-r^2}$ orbital. On the other hand, both $\chi^{0;a}_{1212}(\q)$ and $\chi^{0;a}_{1221}(\q)$ are 
enhanced at $(0.5 \pi, 0.5 \pi)$ 
due to the nesting between the bonding Fermi surface consisting predominantly of $d_{3z^2-r^2}$ orbital and the antibonding Fermi surface having an almost 
equal mixture of both the orbitals although the nesting is relatively weaker than the bonding-bonding case. The longitudinal susceptibility 
is rendered featureless as the peak structure of $\chi^{0;s}_{2222}(\q)$ is subtracted out by the peak structure of $\chi^{0;s}_{1122}(\q)$ which is 
equal to the $\chi^{0;s}_{1221}(\q)$.
\begin{figure}
\begin{center}
\vspace*{-2mm}
\hspace*{0mm}
\psfig{figure=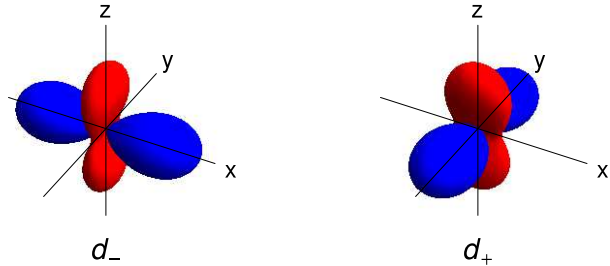,width=100mm,angle=0}
\vspace*{-5mm}
\end{center}
\caption{$d_{\mp}=\frac{1}{\sqrt{2}}(d_{3z^2-r^2}\,{\mp}\,d_{x^2-y^2}$) orbitals with red and blue colors highlighting the phases.}
\label{dens}
\end{figure} 

Fig. 3 shows the longitudinal $\chi_{zz}(\q)$ and the transverse $\chi_{xx}(\q)$ orbital susceptibilities
calculated within the RPA. Transverse antisymmetric susceptibility is enhanced significantly at ($0.5 \pi ,0.5 \pi$) as compared
to the symmetric susceptibility at ($0.6 \pi ,0.6 \pi$) due to the electronic interaction. On the other hand, the longitudinal
susceptibilities are unaffected by the electronic interaction, and are almost featureless. Therefore, the bonding-antibonding band
nesting is responsible for the strong and dominant orbital fluctuations at wavevector ($0.5 \pi ,0.5 \pi$), with 
the nature of fluctuations being antisymmetric and transversal.
In addition, we note that the transversal orbital susceptibility $\chi_{xx}(\q)$ is nothing but the longitudinal 
orbital susceptibility $\chi_{z^{\prime} z^{\prime}}(\q)$ in a new basis consisting of $d_-$ and $d_+$ orbitals,
where $d_{\mp}=\frac{1}{\sqrt{2}}(d_{3z^2-r^2}\,{\mp}\,d_{x^2-y^2}$) (Fig. 4). 
Similarly, the longitudinal susceptibility $\chi_{zz}(\q)$  in the original basis is equal to the transverse 
susceptibility $\chi_{x^{\prime} x^{\prime}}(\q)$. Then, the fact that the longitudinal susceptibilities
$\chi_{x^{\prime} x^{\prime}}(\q)$ (transverse susceptibilities in 
the new basis) are featureless
and the transverse susceptibilities $\chi_{z^{\prime} z^{\prime}}(\q)$ (longitudinal susceptibilities in the basis) show peak structures follows 
from the orbital compositions of the Fermi surface, which consists predominantly of $d_-$ and $d_+$ orbitals along $k_x/\pi$ $\approx \pm$ 0.3 (or $\pm$0.2) and
$k_y/\pi$ $\approx$ $\pm $0.3 (or $\pm$0.2), respectively as shown in Fig. 5.
\begin{figure}
\begin{center}
\vspace*{-2mm}
\hspace*{0mm}
\psfig{figure=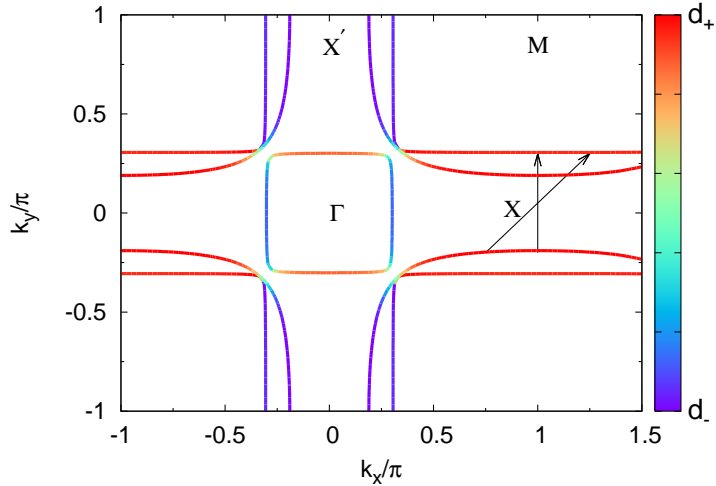,width=70mm,angle=-90}
\vspace*{-5mm}
\end{center}
\caption{Orbital densities of $d_-$ and $d_+$ orbitals on the Fermi surface for the crystal-field parameter $\Delta = -0.3$,  the chemical potentials $\mu = -1.26$, and the inter-planar
 hopping $t_{\perp} = 0.33$.}
\label{dens}
\end{figure} 
\section{Phonon renormalization}
Since strong antisymmetric orbital fluctuations are present due to the correlations, which couples to the antisymmetric component of 
transversal Jahn-Teller phonons, we consider the impact of these orbital fluctuations on the phonon propagator in this section. The second-order perturbation 
theory with respect to the last term of the Hamiltonian given by Eq. (1) provides the following self-energy of transversal Jahn-Teller 
phonons\cite{allen,maier}
\be
\Sigma^{s(a)}(\q, \Omega+i\delta) = - g^2 \sum_{\mu  \nu }\chi^{s(a)}_{\mu \bar{\mu}, \nu \bar{\nu} }(\q, \Omega+i\delta), 
\ee
where 
\be
\chi^{s(a)}_{\mu \bar{\mu}, \nu \bar{\nu}}({\bf q}, \Omega+i\delta) = [\hat{\chi}^{0;s(a)}({\bf q}, \Omega+i\delta)
  [\hat{1}+\hat{U}^{}\hat{\chi}^{0;s(a)}({\bf q}, \Omega+i\delta)]^{-1}]_{\mu \bar{\mu}, \nu \bar{\nu}}.
\ee
\begin{figure}
\begin{center}
\vspace*{-2mm}
\hspace*{0mm}
\psfig{figure=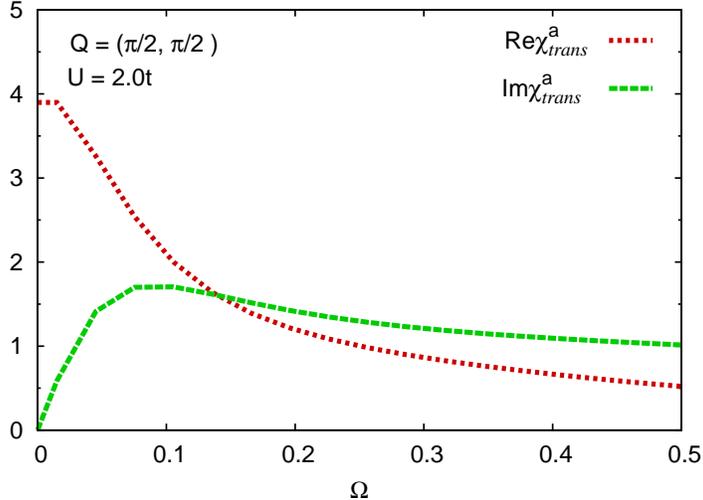,width=70mm,angle=-90}
\vspace*{-5mm}
\end{center}
\caption{Real and imaginary parts of the antisymmetric transverse orbital susceptibility as a function frequency at fixed momentum 
$\q = (0.5 \pi ,0.5 \pi)$.}
\label{self}
\end{figure} 
\begin{figure}
\begin{center}
\vspace*{-2mm}
\hspace*{0mm}
\psfig{figure=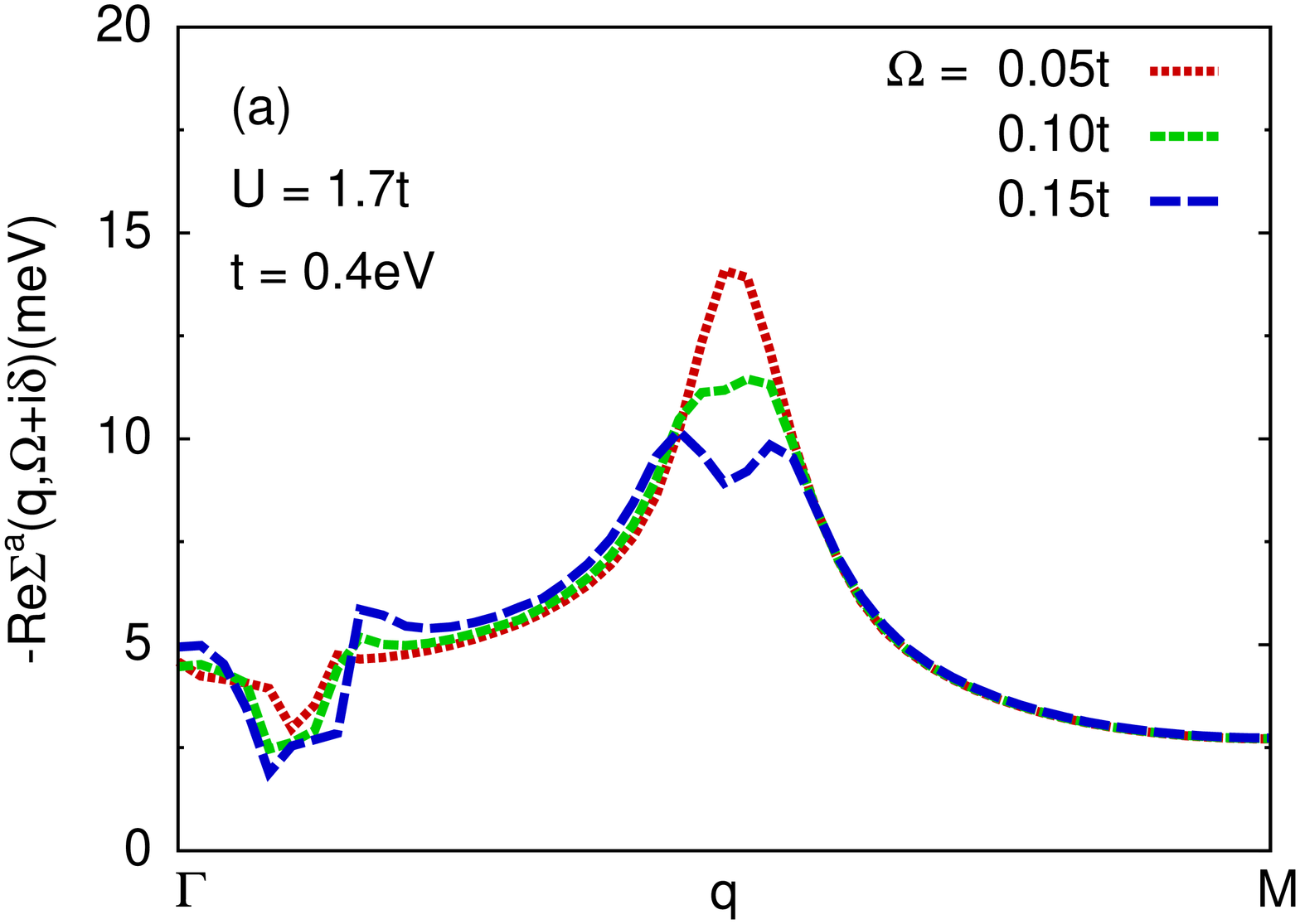,width=55mm,angle=-90}
\psfig{figure=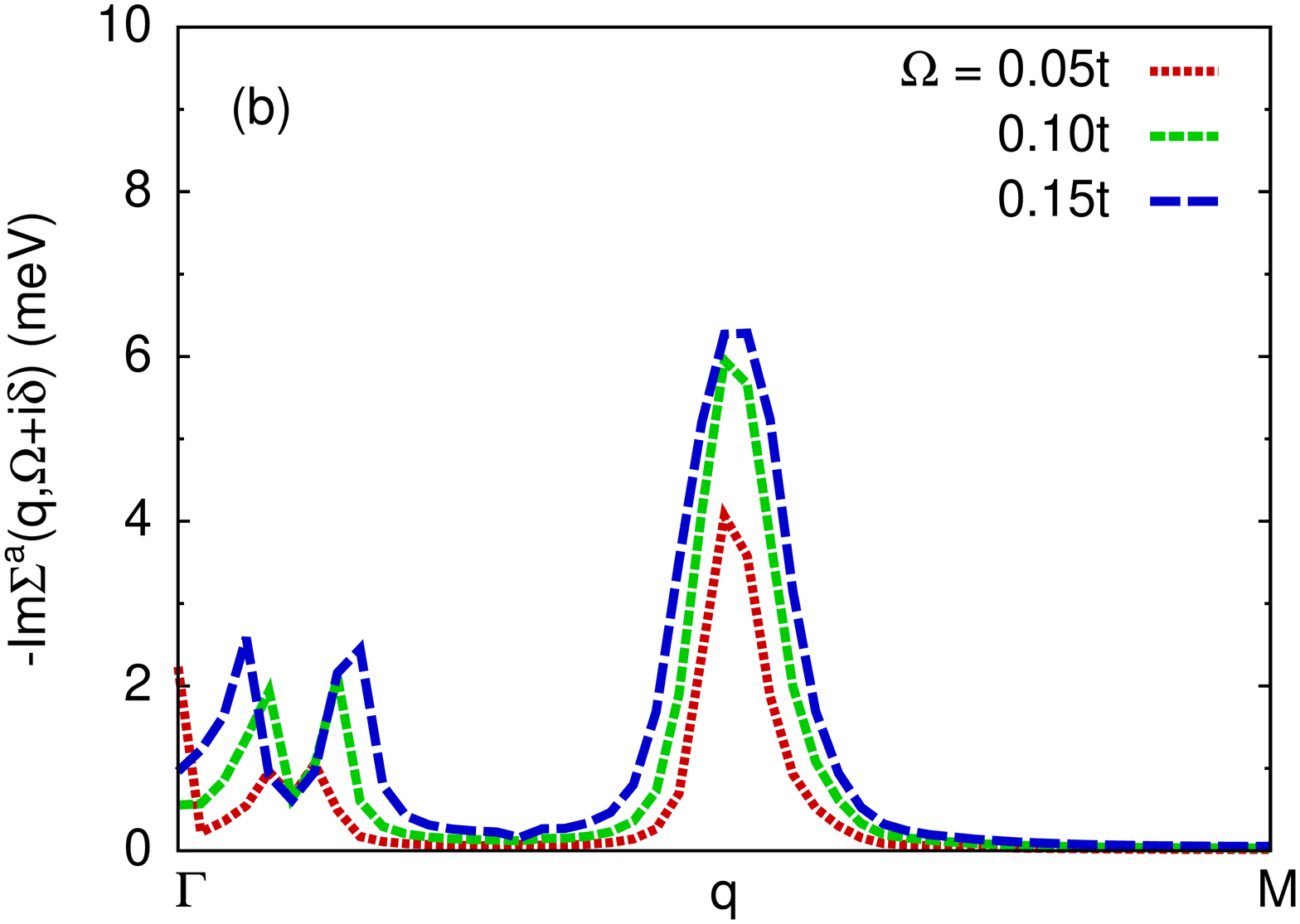,width=55mm,angle=-90}
\vspace*{-5mm}
\end{center}
\caption{(a) Real and (b) imaginary part of the antisymmetric phonon self-energy arising due to 
the antisymmetric transversal orbital fluctuation mode at wavevector (0.5$\pi$, 0.5$\pi$).}
\label{phren}
\end{figure}  

Fig. \ref{self} shows the real and imaginary part of $\chi^a_{trans}(\q, \Omega + i\delta)$ for (0.5$\pi$, 0.5$\pi$) as a function of frequency, where the imaginary part
shows a maximum near $\Omega \approx 0.1t$ which provides the order of energy scale of the fluctuation.
Fig. \ref{phren} shows the real and imaginary part of the phonon self-energy for several values of frequency $\Omega$ around the energy scale of antisymmetric
transverse orbital fluctuation. Here, the frequencies are chosen of the same order as that of local Jahn-Teller
distortion frequency $\Omega^* \approx 50 meV$ determined from Raman scatterings,\cite{allen, iliev} which can be estimated roughly to be $\Omega^* \approx 0.1t$ with $t$
$\approx 0.4 eV$.\cite{dagotto} We have chosen the electron-phonon coupling 
parameter $g \approx 0.125t$ to reproduce the experimental result as shown in Fig. 7. There is a large self-energy correction to antisymmetric component of transversal Jahn-Teller phonons 
near (0.5$\pi$, 0.5$\pi$, 0) where the linewidth also exhibits maximum. Both these features show a good quantitative agreement  
with the neutron scattering experiments for phonons in the ferromagnetic-metallic bilayers.\cite{weber}

We also find additional 
peaks in the the linewidth and dip in the self-energy for the low-momentum region. Already existing on either side of 
$|\q|\approx 0.1 \pi$ at the bare level, these peaks appear due to the saddle point behavior of $E^a_{-}({\bf k+q})-E^s_{-}({\bf k})$ 
and $E^s_{-}({\bf k+q})-E^a_{-}({\bf k})$ in the denominator of the antisymmetric susceptibility for  $|\q| \ge 0.1 \pi$ or 
$|\q| \le 0.1 \pi$, respectively. Saddle points lie near $\k$ = ($-0.3 \pi, \mp \pi$) and ($\mp \pi, -0.3 \pi$) with maximum along $k_x$ and minimum along $k_y$ 
for $E^a_{-}({\bf k+q})-E^s_{-}({\bf k})$, and near $\k$ = $(0.3 \pi, \mp \pi)$ and $(\mp \pi, 0.3 \pi)$ with minimum along
$k_x$ and maximum along $k_y$ for $E^s_{-}({\bf k+q})-E^a_{-}({\bf k})$. The movement in the opposite directions and enhancement of the peaks away from $|\q|\approx 0.1 \pi$ with 
increasing energy follows from the movement towards the saddle point which lies only slightly away from the Fermi surface in each case.
The origin of these peak structures, therefore, is analogous to that of peak found in the density of 
states near the saddle point of the energy band in the momentum space. Thus, our study suggests additional peaks to 
be observed in the inelastic neutron scattering experiments in low-momentum region.
\begin{figure}
\begin{center}
\vspace*{-2mm}
\hspace*{0mm}
\psfig{figure=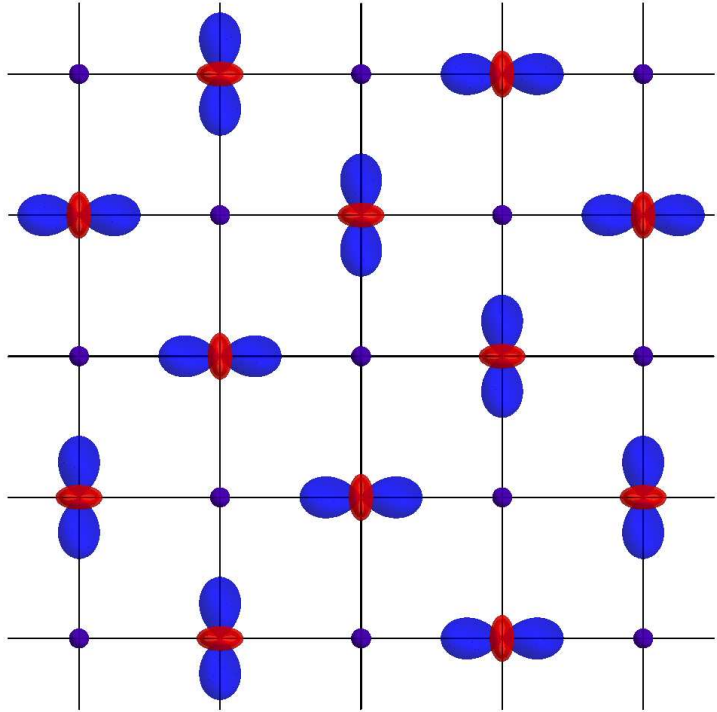,width=100mm,angle=0}
\vspace*{-5mm}
\end{center}
\caption{Orbital correlations with wavevector (0.5$\pi$, 0.5$\pi$) developed in the ferromagnetic metallic La$_{1.2}$Sr$_{1.8}$Mn$_2$O$_7$ involving $d_{\mp}$ orbitals, where electrons hop along zigzag chains
in $x$ and $y$ directions through $d_-$ and $d_+$ orbitals, an essential mechanism to stabilize the CE-type spin arrangement.}
\label{dens}
\end{figure} 
\vspace*{1\baselineskip}
\section{Conclusions and Discussions}
Our investigation of the orbital fluctuations in a bilayer system with two orbitals at each site, which captures the salient features of the ARPES measurements
 for the ferromagnetic-metallic phase of the doped bilayer manganite La$_{1.2}$Sr$_{1.8}$Mn$_2$O$_7$ has provided an important connection between 
 the the Fermi-surface structure and the short-range dynamical correlations as observed in the neutron scattering experiments. 
 Antisymmetric longitudinal orbital fluctuations in the basis of  $d_-$ and $d_+$ orbitals are strong at
($ 0.5 \pi, 0.5 \pi$), where $d_{\mp}=\frac{1}{\sqrt{2}}(d_{3z^2-r^2}\,{\mp}\,d_{x^2-y^2})$. This follows from the Fermi-surface nesting
between bonding and antibonding bands instead of the nesting between bonding and bonding bands despite their
flat segments. Moreover, the orbital fluctuations strongly renormalize 
the antisymmetric component of transversal Jahn-Teller phonons near wavevector (0.5$\pi$, 0.5$\pi$), and are responsible for the shortest lifetime of the phonons near the same wavevector, 
implying their dynamic nature. Our study also predicts the enhancement of phonon linewidth near the low momentum region, which should be 
observed in the inelastic neutron scattering experiments.

The proximity to the orbital 
ordering instability of La$_{1.2}$Sr$_{1.8}$Mn$_2$O$_7$ due to the strong nesting with orbital 
compositions of the Fermi surface predominantly of $d_-$ and $d_+$ orbitals along $k_x/\pi$ $\approx \pm$ 0.3 (or $\pm$0.2) and
$k_y/\pi$ $\approx$ $\pm $0.3 (or $\pm$0.2) is suppressed by the ferromagnetism stabilized 
by the double-exchange mechanism.\cite{zener} In such orbitally correlated or ordered state, the electrons will prefer
to hop along $x$ and $y$ directions through $d_-$ and $d_+$ orbitals as shown in Fig. 8, which may instead support a CE-type spin 
arrangement wherein spins are aligned ferromagnetically along the zigzag chains. Therefore, our 
study provides a plausible explanation for the existence
of CE-type orbital fluctuations in the ferromagnetic-metallic bilayer near $x = 0.4$ as observed in the neutron scattering experiments.
\section* {Acknowledgements}

This work is supported by Basic Science Program through the National Research
Foundation of Korea (NRF) funded by the Ministry of Education (NRF-2012R1A1A2008559). 
D. K. Singh would also like to acknowledge the Korea Ministry of Education, Science and Technology, Gyeongsangbuk-Do and
Pohang City for the support of Young Scientist Training Program  at the Asia Pacific Centre for Theoretical Physics. K. H. Lee would also
like to thank the Korea Ministry of Education, Science and Technology, Gyeongsangbuk-Do and
Pohang City for the support of Independent Junior Research Groups at the Asia Pacific
Centre for Theoretical Physics and NRF-2012R1A1A2008028.


\begin{thebibliography}{08}

\bibitem{helmolt}
R. von Helmolt, J. Wecker, B. Holzapfe, L. Schultz, and K. Samwer, 
\phrl {\bf 71}, 2331 (1993).

\bibitem{jin}
S. Jin, T. H. Tiefel, M. McCormack, R. A. Fastnacht, R. Ramesh, and L. H. Chen, 
Science {\bf 264}, 413 (2004).
 
\bibitem{sen}
C. \c{S}en, G. Alvarez, and E. Dagotto, Phys. Rev. Lett. {\bf 105}, 097203 (2010).

\bibitem{weber}
F. Weber, N. Aliouane, H. Zheng, J. F. Mitchell, D. N. Argyriou, and D. Reznik, Nature Mater. {\bf 8}, 798 (2009).

\bibitem{saitoh}
E Saitoh, Y Tomioka, T Kimura, and Y Tokura, J. Magn.
Magn. Mater. {\bf 239}, 170 (2002).


\bibitem{adams}
C. P. Adams, J.W. Lynn, Y. M. Mukovskii, A. A. Arsenov, and D. A. Shulyatev,
 Phys. Rev. Lett. {\bf 85}, 3954 (2000).

\bibitem{moussa_2007}
F. Moussa, M. Hennion, P. Kober-Lehouelleur, D. Reznik, S. Petit, H. Moudden, A. Ivanov, Ya. M. Mukovskii, R. Privezentsev, and F. Albenque-Rullier,
Phys. Rev. B {\bf 76}, 064403 (2007).

\bibitem{wollan_1955}
E. O. Wollan and W. C. Koehler,
Phys. Rev. {\bf 100}, 545 (1955).

\bibitem{goodenough_1955}
J. B. Goodenough,
Phys. Rev. {\bf 100}, 564 (1955).

\bibitem{hwang} 
H. Y. Hwang, P. Dai, S-W. Cheong, G. Aeppli, D. A. Tennant, and
H. A. Mook, Phys. Rev. Lett. {\bf 80}, 1316 (1998).
 
\bibitem{ye} 
F. Ye, P. Dai, J. A. Fernandez-Baca, H. Sha, J.W. Lynn, H. Kawano-
Furukawa, Y. Tomioka, Y. Tokura, and J. Zhang, Phys. Rev. Lett.
{\bf 96}, 047204 (2006).

\bibitem{dk1}
D. K. Singh, B. Kamble, and A. Singh, Phys. Rev. B {\bf 81}, 064430
(2010).

\bibitem{dk2}
D. K. Singh, B. Kamble, and A. Singh, J. Phys.: Condens. Matter
{\bf 22}, 396001 (2010).

\bibitem{moritomo}
Y. Moritomo, A. Asamitsu, H. Kuwahara, and Y. Tokura, Nature
{\bf 380}, 141 (1996).

\bibitem{mitchell}
 J. F. Mitchell, D. N. Argyriou, J. D. Jorgensen, D. G. Hinks, C. D. Potter, and S. D. Bader,
 Phys. Rev. B {\bf 55}, 63-6 (1997).

\bibitem{chuang}
Y.-D. Chuang, A. D. Gromko, D. S. Dessau, T. Kimura, and 
Y. Tokura, Science {\bf 292}, 1509 (2001).

\bibitem{mannela}
N. Mannella, W. L. Yang, X. J. Zhou, H. Zheng, J. F. Mitchell, J. Zaanen, T. P. Devereaux,
N. Nagaosa, Z. Hussain, and Z.-X. Shen,
 Nature {\bf 438}, 474 (2005).

\bibitem{evtushinsky}
D.V. Evtushinsky, D. S. Inosov, G. Urbanik, V. B. Zabolotnyy, R. Schuster, P. Sass, T. H\"{a}nke, C. Hess,
B. B\"{u}chner, R. Follath, P. Reutler, A. Revcolevschi, A. A. Kordyuk, and S.V. Borisenko,
\phrl{105}, 147201 (2010).

\bibitem{zsun1}
Z. Sun, Y.-D. Chuang, A.V. Fedorov, J. F. Douglas, D. Reznik, F.Weber, N. Aliouane, D. N. Argyriou, H. Zheng,
J. F. Mitchell, T. Kimura, Y. Tokura, A. Revcolevschi, and D. S. Dessau, Phys. Rev. Lett. {\bf 97}, 056401 (2006).

\bibitem{jong}
S. de Jong, F. Massee, Y. Huang, M. Gorgoi, F. Schaefers, J. Fink, A. T. Boothroyd,
D. Prabhakaran, J. B. Goedkoop, and M. S. Golden, Phys. Rev. B {\bf 80}, 205108 (2009).

\bibitem{zsun2}
Z. Sun, J. F. Douglas, A. V. Fedorov, Y.-D. Chuang, H. Zheng, J. F. Mitchell, and D. S. Dessau, 
Nature Phys. {\bf 3}, 248 (2007).


\bibitem{masse}
F. Massee, S. de Jong, Y. Huang, W. K. Siu, I. Santoso, A. Mans, A. T. Boothroyd,
D. Prabhakaran, R. Follath, A. Varykhalov, L. Patthey, M. Shi, J. B. Goedkoop, and M. S. Golden, 
Nature Phys. {\bf 7}, 978 (2011).

\bibitem{zsun3}
Z. Sun, Q. Wang, J. F. Douglas, Y.-D. Chuang, A. V. Fedorov, E. Rotenberg, H. Lin, S. Sahrakorpi, B. Barbiellini, 
R. S. Markiewicz, A. Bansil, H. Zheng, J. F. Mitchell, and D. S. Dessau, 
\phrb{86}, 201103 (2012).

\bibitem{zsun4}
Z. Sun, Y.-D. Chuang, A.V. Fedorov, J. F. Douglas, D. Reznik, F.Weber, N. Aliouane, D. N. Argyriou,
H. Zheng, J. F. Mitchell, T. Kimura, Y. Tokura, A. Revcolevschi, and D. S. Dessau, 
Phys. Rev. Lett. {\bf 97}, 056401 (2006).

\bibitem{dagotto}
E. Dagotto, T. Hotta, and A. Moreo,  Phys. Rep. {\bf 344} 1 (2001).

\bibitem{eremin}
I. Eremin, D. K. Morr, A. V. Chubukov, and K. Bennemann, 
Phys. Rev. B {\bf 75}, 184534 (2007).


\bibitem{saniz}
R. Saniz, M. R. Norman, and A. J. Freeman, \phrl {\bf 101}, 236402 (2008).



\bibitem{takimoto}
T. Takimoto, T. Hotta, T. Maehira, and K. Ueda, 
J. Phys.: Condens. Matter {\bf 14}, L369-L375 (2002).


\bibitem{allen}
P. B. Allen, 
Phys. Rev. B {\bf 6}, 2577 (1972).

\bibitem{maier}
T. A. Maier, S. Graser, P. J. Hirschfeld, and D. J. Scalapino, 
Phys. Rev. B {\bf 83}, 220505 (2011). 

\bibitem{allen}
V. Perebeinos and P. B. Allen,  
\phrl {\bf 85}, 5178 (2000).

\bibitem{iliev}
M. N. Iliev, M. V. Abrashev, H.-G. Lee, V. N. Popov,
Y. Y. Sun, C. Thomsen, R. L. Meng, and C. W. Chu, Phys.
Rev. B {\bf 57}, 2872 (1998).

\bibitem{zener}
C. Zener, Phys. Rev. {\bf 82}, 403 (1951).
\end{thebibliography}
\end{document}